\documentclass{elsart}

\usepackage{graphics}

\usepackage{graphicx}

\usepackage{epsfig}

\usepackage{amssymb}

\begin{document}

\begin{frontmatter}

\title{Anisotropy parameters of  superconducting MgB$_2$}

\author{V.G. Kogan\corauthref{cor1},}
\corauth[cor1]{Corresponding author.}
\ead{kogan@ameslab.gov}
\author{S.L. Bud'ko}
\address{Ames Laboratory and Department of Physics and Astronomy, Iowa
State University, Ames, IA 50011, USA}

\begin{abstract}
Data on   macroscopic superconducting anisotropy
of MgB$_2$ are reviewed. The data are   described
within a weak coupling  two-gaps anisotropic s-wave model of
superconductivity. The calculated ratio of the upper critical
fields $\gamma_H=H_{c2,ab}/H_{c2,c}$ increases with decreasing
temperature in agreement with available data, whereas the calculated
ratio of London penetration depths
$\gamma_{\lambda}=\lambda_c/\lambda_{ab}$ decreases to reach
$\approx 1.1$ at $T=0$. Possible macroscopic consequences of
$\gamma_{\lambda}\ne\gamma_H$ are discussed.
\end{abstract}

\end{frontmatter}

\section{Introduction}

  Physics and applications of new superconducting
materials cannot be properly understood and developed without careful
characterization of possible anisotropies. High-$T_c$
materials are one of the best examples of  relevance of the
superconducting anisotropy to all aspects of physics and possible
applications.

It is a common practice to characterize anisotropic
superconductors by a {\it single} anisotropy parameter defined as
$\gamma=\xi_a/\xi_c\equiv\lambda_c/\lambda_a$. Here, $\xi$ is the coherence
length, $\lambda$ is the penetration depth, and $a,c$ are principal
directions of a uniaxial crystal of the interest here. Often
$\xi_a=\xi_b$ are denoted in literature as $\xi_{ab}$; for formal
reasons we   prefer a single subscript notation. Since the upper critical
fields, $H_{c2,c}=\phi_0/2\pi\xi_{a }^2$ and
$H_{c2,a }=\phi_0/2\pi\xi_{a } \xi_c$, the anisotropy parameter can
also be written as $\gamma=H_{c2,a }/H_{c2,c}$. Historically, the
practice to assign a single anisotropy parameter to each material
emerged after the anisotropic Ginzburg-Landau (GL) equations were
proposed phenomenologically by Ginzburg
\cite{Ginz} and derived microscopically for  arbitrary gap and Fermi
surface anisotropies by
   Gor'kov and  Melik-Barkhudarov \cite{Gorkov}.   Formally, this came
out because near the critical temperature $T_c$, the same ``mass
tensor" $m_{ik}$  determines the anisotropy  of both $\xi$ (of the
upper critical fields $H_{c2}$) and of $\lambda$:
\begin{equation}
(\xi^2)_{ik} \propto m_{ik}^{-1}\,,\qquad (\lambda^2)_{ik} \propto
m_{ik}\,.
\end{equation}
Therefore,
\begin{equation}
\gamma_{\lambda}^2={\lambda^2_c\over
\lambda^2_{a}}=\frac{m_{cc}}{m_{aa}}=
\frac{\xi^2_{a }}{\xi^2_{c}}=\gamma^2_{\xi}   \,.
\end{equation}

At arbitrary temperatures, however, the theoretical approach  for
calculating $H_{c2}$ (the position of the second order phase
transition in high fields) has little in common with evaluation of
$\lambda$ (the weak-field relation between the
current and the vector potential), so that the anisotropies of these
quantities  are not necessarily the same. In fact, in materials with
anisotropic Fermi surfaces and anisotropic gaps, not only the
parameter $\gamma_H=H_{c2,a}/H_{c2,c}$ may strongly depend on
$T$, but this ratio might differ considerably from
$\gamma_{\lambda}=\lambda_c/\lambda_a$ at low temperatures
\cite{KRap,MMK}.

In this brief review we  collect  data on anisotropy of MgB$_2$
available to us, discuss briefly the methods used and the results
obtained. Then we outline the weak-coupling microscopic approach which
can be used to evaluate the anisotropy parameters and their $T$
dependence.  We conclude with discussion of  macroscopic
consequences of different $H_{c2}$ and $\lambda$ anisotropies   focussing
on the torque problem, the quantity commonly measured to extract the
anisotropy parameters.

\section{Data review}

   Experimentally, superconducting anisotropy of MgB$_2$ was
under scrutiny right away after the discovery of this material. Almost
all experimental studies were concerned with the
$H_{c2}$ anisotropy, $\gamma_H = H_{c2,ab}/H_{c2,c}$.

Based on the AC susceptibility and the field dependent magnetization
measurements on sorted powders, magnesium diboride was
claimed to be an isotropic superconductor \cite{Chen22}. The $H_{c2}$
anisotropy for separate particles settled on a flat surface was reported
as $\gamma_H\approx 1.73$ \cite{deLima}. For a hot-pressed
bulk samples  values of $\gamma_H$ up to 1.2 were obtained
  from susceptibility and resistance measurements \cite{Handstein}.
Magnetotransport measurements on MgB$_2$ thin films
with different degrees of $c$-axis orientation yielded $\gamma_H =
1.8-2$  ($T_c$ = 31-37 K) with higher anisotropy for films of higher
resistivity \cite{Patnaik}. The temperature-independent $\gamma_H$ = 1.2
  ($T_c$ = 39 K) was reported for $c$-oriented films, Ref.
\cite{Mia}. Similar results ($\gamma_H$ = 1.2 - 1.8) were later
obtained for a set of three films pulsed laser deposited (PLD) on
different substrates  ($T_c$ = 31-37 K) \cite{Ita1,Ita2}.  In this case
no apparent correlation was observed between the $H_{c2}$ anisotropy and
residual resistance ratio (RRR) or superconducting transition temperature
of the films. Quite different values, $\gamma_H$ = 9 - 13, were reported
for {\it in-situ} grown PLD films with $T_c$ suppressed down to 24 - 27 K
\cite{Shinde}.

In a different approach, the $H_{c2}$ anisotropy of magnesium diboride
was evaluated from a number of  measurements on randomly
oriented powders. Analysis of the conduction electron spin resonance (CESR)
   data on fine MgB$_2$ powders taken in a wide temperature
range  yielded an estimate of $\gamma_H \approx$ 8 \cite{Simon}.
Similar idea (deconvoluting the signal in a certain temperature
range in  two components,
one corresponding to superconducting state and the other originating from
the normal state) was later used for interpretation of the $^{11}$B
nuclear magnetic resonance (NMR) data in the mixed
state of MgB$_2$ \cite{Papavassi}. These measurements resulted in
   $\gamma_H \approx$ 6. Subsequent
$^{11}$B NMR and magnetization studies on Mg$_{1-x}$Al$_x$B$_2$
($x \le$ 0.025)  reveal a considerable decrease of
the   anisotropy with Al doping \cite{Pissas}.

In addition to local NMR and CESR probes, global measurements on samples
with randomly oriented grains were   used for the
anisotropy studies. High field reversible DC magnetization data  were well
described  assuming that MgB$_2$ is an anisotropic superconductor with
$\gamma_H$ = 6 - 9 and using the available experimental estimates of the
average penetration depth
$\lambda (0) = (\lambda_{ab}^2\lambda_c)^{1/3}  =
110-140$ nm \cite{Simon}. The $H_{c2}$ anisotropy was also
evaluated from the broadening of the resistive transition in  applied
magnetic field. Two slightly different theoretical approaches to this
problem, \cite{Glazman} and \cite{Welch},   based on the
anisotropic Ginzburg-Landau and percolation theories
for randomly oriented anisotropic   superconducting grains,    were
developed more than a decade ago in early days of the high  $T_c$
superconductivity. They require somewhat different analysis of the
resistivity data but result in   similar estimates: $\gamma_H$ = 5 - 7
\cite{Madison}.

\begin{figure}
\begin{center}
\includegraphics*[width=12cm]{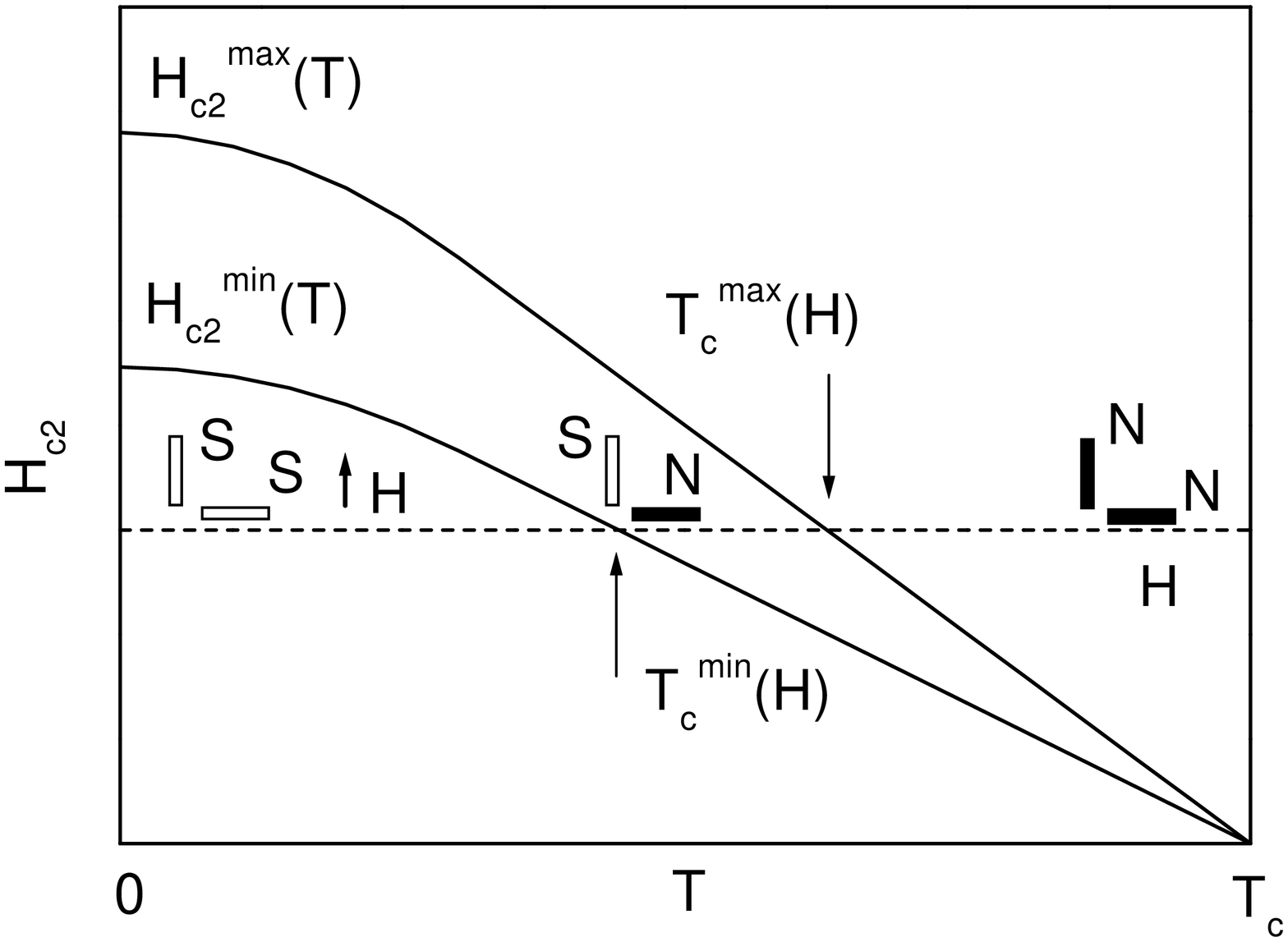}
\end{center}
\caption{Sketch of the maximum  $H_{c2,max}(T)\equiv H_{c2,a}$ and the
minimum
$H_{c2,min}(T)\equiv H_{c2,c}$ upper critical fields. For a given applied
field
$H$, the relation
$H = H_{c2,min}(T_c^{min}) = H_{c2,max}(T_c^{max})$  defines
temperatures $T_c^{min},T_c^{max}$. The open (shaded) rectangles
represent superconducting (normal) grains.}
\label{sketch}
\end{figure}

The aforementioned determinations of the $H_{c2}$ anisotropy utilized
   traditional techniques, with the exception of those based on
CESR and NMR. About a year ago a robust and a simple way to extract this
anisotropy  from the data on $T$  dependence of the magnetization of {\it
random powders} taken in various fixed fields
has been suggested \cite{BKC}. The method is based on two features in
$(\partial M/\partial T)_H$:
the onset of diamagnetism at $T_c^{max}$, which is
commonly  associated with $H_{c2}$ and which is, in fact,
$H_{c2,max}\equiv H_{c2,a}$, and a  kink in
$\partial M/\partial T$ at a lower temperature $T_c^{min}$ (see Fig.
\ref{sketch}). The origin of these two features can be understood as
follows.

Upon cooling a powder sample in a fixed $H$, there is  a deviation
from a roughly $T$ independent normal state magnetization
(for non-magnetic materials as is the case of MgB$_2$) to an
increasingly diamagnetic signal at $T= T_c^{max}$. On the $H-T$ diagram,
the point $(H,T_c^{max})$ lies at the phase boundary $H_{c2,max}(T)$.
Repeating this measurement at various fixed field one recovers the full
curve $H_{c2,max}(T)\equiv H_{c2,a}(T)$. This line   coincides with the
upper critical field determined for polycrystalline samples by, e.g., the
standard resistivity measurements. A second sharp
feature   occurs in $\partial M/\partial T$ when the sample
temperature passes through $T_c^{min}$.  This can be readily
understood by considering what happens to the sample upon warming.
For $T < T_c^{min}$ all randomly oriented grains are superconducting,
whereas for $T > T_c^{min}$ part of them are normal,
depending upon their orientation with respect to the applied magnetic
field.  Therefore, upon warming through $T_c^{min}$ we expect some
peculiarity, or a kink in $\partial M/\partial T$ associated with the onset
of normal state properties in an increasing number of appropriately
oriented grains.  In a similar spirit the field dependent magnetization
can be analyzed. This method constitutes a robust
procedure independent either of a particular model for describing the
anisotropy or of a degree of powder randomness. A  kink in
$\partial M/\partial T$  at $T = T_c^{min}$  should be
present for any angular distribution of the grains (as long as it is
continuous, although not necessarily random).

It should be noted that
in superconductors with strong fluctuations (like HTSC) or in the samples
with a distribution of $T_c$'s or other physical properties due to
chemical or/and structural inhomogeneities
the region of superconductor-normal transition and the kink in
$\partial M/\partial T$ may be smeared. In such cases this simple method
may not work. Still, given difficulties in growing single crystals of
a sufficient size often encountered when a new material is synthesized, the
method may prove useful and sometimes the only one available for
determination of the $H_{c2}$ anisotropy.

\begin{figure}
\begin{center}
\includegraphics*[width=12cm]{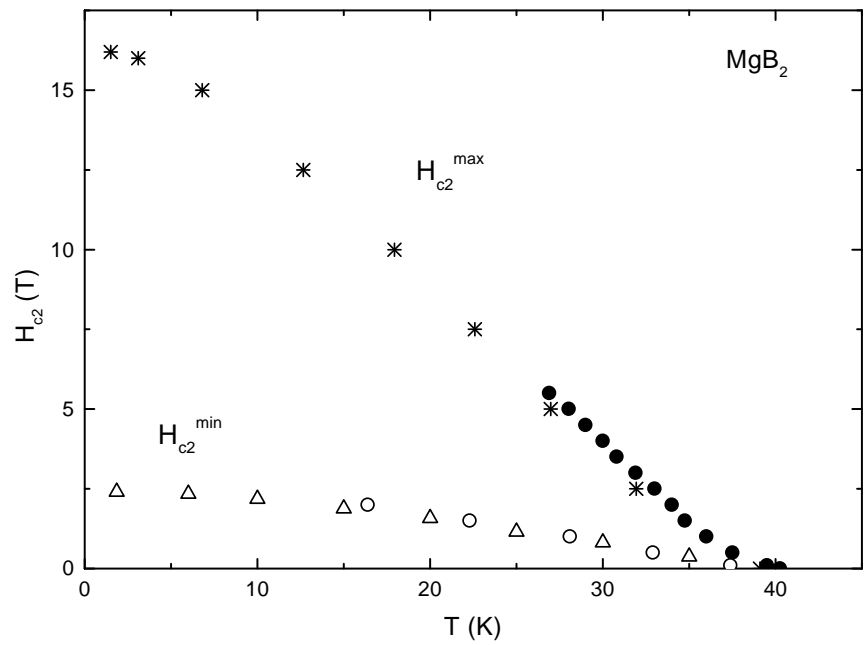}
\end{center}
\caption{The $H-T$ phase diagram for anisotropic MgB$_2$. Symbols:
circles (open and filled) are extracted  from $M(T)|_H$,
triangles - from $M(H)|_T$,
asterisks - from resistivity data on polycrystalline MgB$_2$.}
\label{Hc2T}
\end{figure}

   The analysis of detailed temperature and field dependent
magnetization measurements on MgB$_2$ powders together with the data on
polycrystalline $H_{c2}$ ($= H_{c2}^{max}$) obtained from high field
magnetotransport measurements on similar samples yielded a complete
anisotropic $H_{c2}(T)$ phase diagram  \cite{BC}, see Fig. \ref{Hc2T}.

A number of direct measurements  of the $H_{c2}$ anisotropy on single
crystals was reported. Magnetoresistance data on small MgB$_2$
single crystalline platelets \cite{Xtal1,Xtal2,Pradhan,Xtal3} gave
$\gamma_H$ =  2.6 - 3.0 (in these samples $T_c$ = 38.0-38.6 K, $RRR$ = 5 -
8). It is argued in Ref. \cite{solo}  that electrical
transport measurements are not well suited to probe bulk upper critical
field of MgB$_2$ due to possible presence of a second phase (related to
surface effects) with enhanced superconducting parameters. In this paper,
the bulk
$H_{c2}^{ab}(T)$ and $H_{c2}^c(T)$ are evaluated utilizing data on
the in-plane thermal conductivity  for two orientations of
the applied magnetic field. The extrapolated value of $\gamma_H$ is
  $\gamma_H(T \rightarrow 0)$ = 4.2 \cite{solo}.

The torque magnetometry was also employed to study
superconducting anisotropy on single crystals \cite{angst}. From angular
dependent torque measurements taken in different applied fields the
anisotropy of the upper critical field was found to be temperature
dependent: it decreases  from $\gamma_H \approx$ 6 at 15 K to $\approx$
2.8 at 35 K. Reversible angular dependent torque measurements (with a
vortex shaking process \cite {Willemin}) performed in the temperature
range 27 - 36 K and in   fields up to 10 kOe were analyzed using a
formula  derived in
\cite{K88} for $\gamma_H=\gamma_{\lambda}$, see Eq. (\ref{trq_0}) below.
As a result, nearly linear field dependence of $\gamma$ has been extracted
with $\gamma (H \rightarrow 0) \approx$ 2 and $\gamma $(10 kOe)
$\approx$ 3.7 with practically no $T$ dependence between
27 K and 36 K.
Extensive torque data of Ref. \cite{torque1} were also interpreted
assuming $\gamma_H=\gamma_{\lambda}=\gamma$ and produced estimates
$\gamma\approx 3 - 4$ which vary  somewhat with $T$ and $H$.

\begin{figure}
\begin{center}
\includegraphics*[width=12cm]{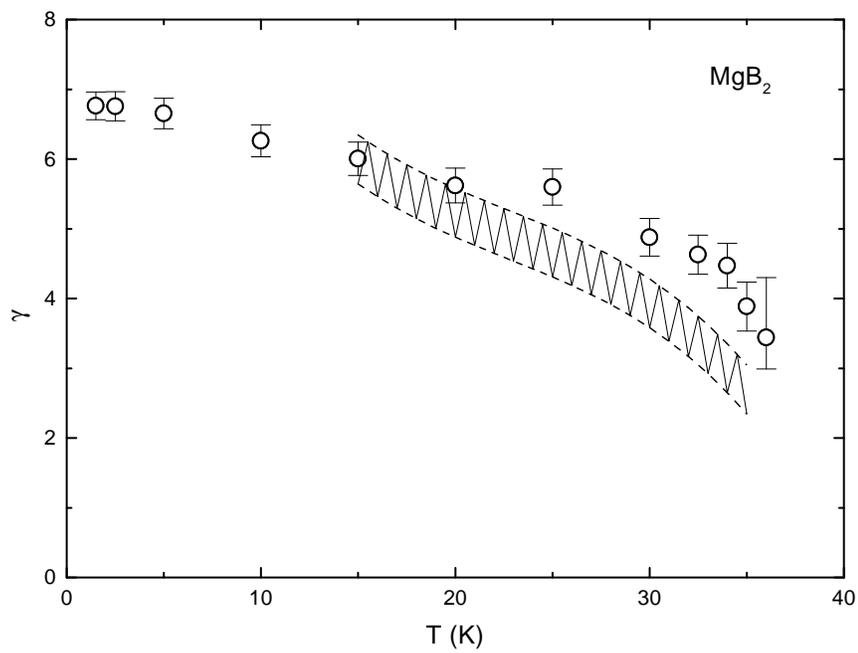}
\end{center}
\caption{Temperature-dependent anisotropy of the upper critical field.
The range of data from Ref. \cite{angst} is shown  for comparison as
a hatched area between  dashed lines.}
\label{gammaT}
\end{figure}

Subsequent analysis \cite{BC} of
the anisotropic $H_{c2}(T)$ data obtained on random MgB$_2$ powders (Fig.
\ref{Hc2T}) showed $\gamma_H (T)$ dependence consistent (Fig.
\ref{gammaT}) with the one observed in the single crystals \cite {angst}.
Similar temperature-dependent
$\gamma_H$ behavior was extracted from magnetotransport, magnetization,
ac susceptibility and specific heat in applied magnetic field measurements on
single-crystalline platelets \cite{Welp,Lyard,Perkins} and in
thin films \cite{Ita2}.

Summarizing,   the reported anisotropy of the upper critical field
in magnesium diboride range from $\gamma_H \approx$ 1 to $\gamma_H
\approx$ 13. There are some concerns related to the spread of these
results such as purity and homogeneity of materials used, degree of
alignment of crystallites/grains, possible sharp angular
dependence of $H_{c2} $ so that correct evaluation of the anisotropy
may suffer from even slight misorientations   in
measurements on single crystallites, and, last but not least, adequacy of
models used to analyze the data on polycrystals and crystals. Even for
similar small single crystals different apparently direct measurements give
somewhat different estimates of $\gamma_H$ (see e.g. \cite{angst,solo}).
On the other hand,   there is an apparent consistence between recent
data on the temperature dependence of $\gamma_H$ obtained by at least two
groups on single crystals  and on high quality powders, similar
"convergence" emerges in  data for other physical properties. This
indicates that we are coming closer to measuring the {\it intrinsic}
properties of MgB$_2$. As a side remark it should be mentioned that
temperature dependent $H_{c2}$ anisotropy is not unique for MgB$_2$,
it was observed, for example, in
NbSe$_2$ \cite{NbSe2}.

For sub-mm size MgB$_2$ single crystal, an anisotropic
lower critical field, $H_{c1}$, was estimated
from the $M(H)$ data \cite{Xtal1}. The $H_{c1}^{ab}(T)$
dependence was reported as near linear with $H_{c1}^{ab}(0) \approx$ 384 G,
while the $H_{c1}^c(T)$ appeared to be non-linear with
and $H_{c1}^c(0)
\approx$ 272 G. Another estimate of the temperature dependent
anisotropic $H_{c1}$ from magnetization measurements on MgB$_2$ single
crystallites \cite{Perkins} gave close to linear $H_{c1}(T)$ for
both directions with extrapolated to $T =$ 0 values
$H_{c1}^c(0) \approx$ 2800 G and $H_{c1}^{ab}(0) \approx$ 1300 G
and near $T$ independent ratio
$H_{c1}^c/H_{c1}^{ab} \approx$ 2.2.

Apparently more work  on sample quality, measurement procedures,
measurements analysis, and modelling is required to obtain a better
physical picture of the superconducting anisotropy of MgB$_2$. Reliable
data on the  temperature dependence of the anisotropy   of the
London penetration depth, when available, will serve as an important test
of existing models.

\section{Theory of anisotropies of $\bf\lambda$ and $\bf H_{c2}$}

To address theoretically the question of anisotropy at arbitrary
temperatures one has to turn to a microscopic model. The most
complete and sophisticated approach is based on taking into account
the actual Fermi surface, the phonon distribution, and the
electron-phonon interaction (all of which are anisotropic) in the frame
of the Eliashberg theory. This can be and has been done numerically by
a few groups, which evaluated the Fermi
surfaces \cite{Mazin,Pikett} and the gap distribution
\cite{Choi,Mazin1}.

\begin{figure}
\begin{center}
\includegraphics*[width=12cm]{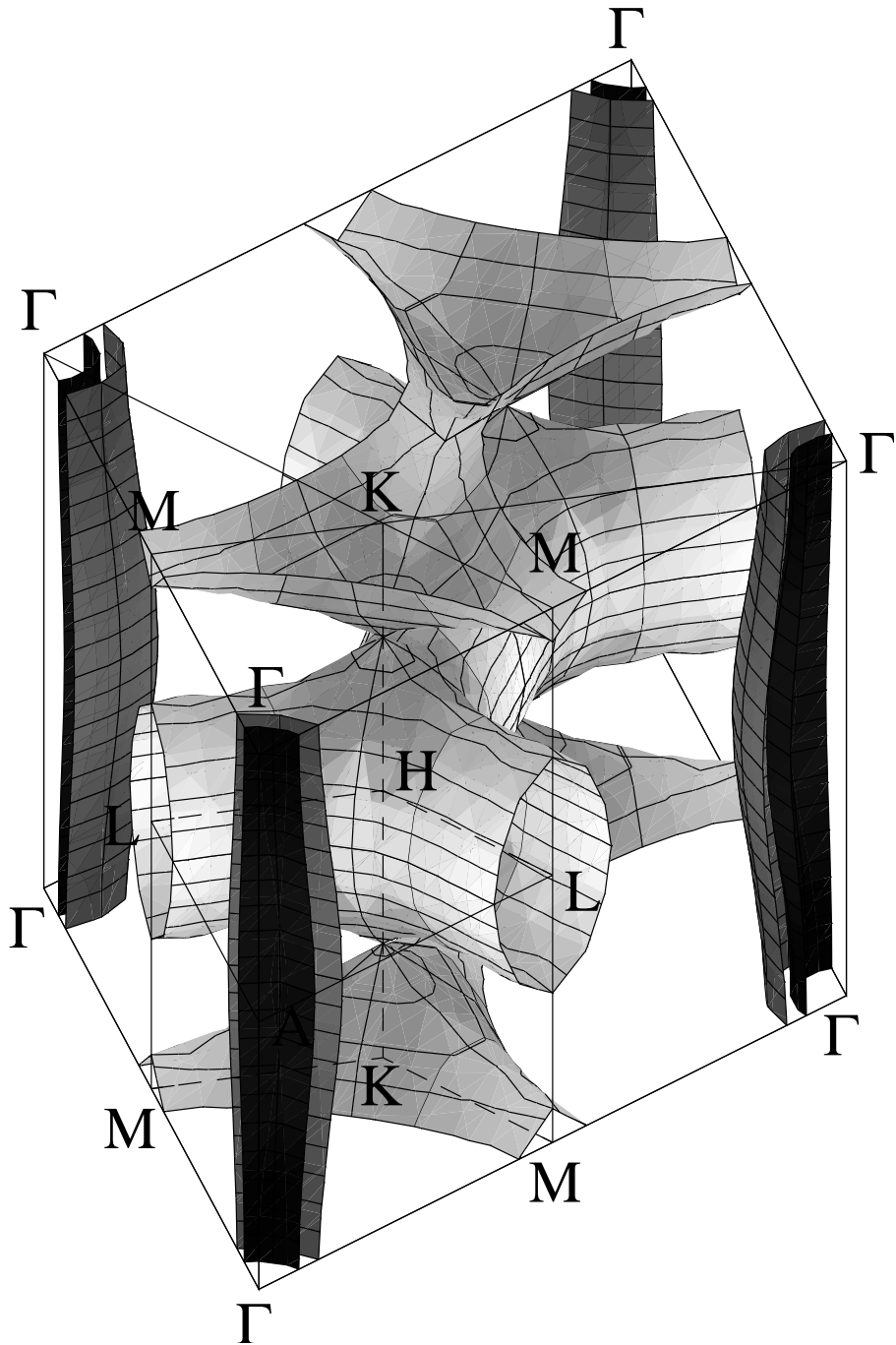}
\end{center}
\caption{Fermi surface of MgB$_2$ calculated in Ref. \cite{Choi}.}
\label{Choi1}
\end{figure}

\begin{figure}
\begin{center}
\includegraphics*[width=12cm]{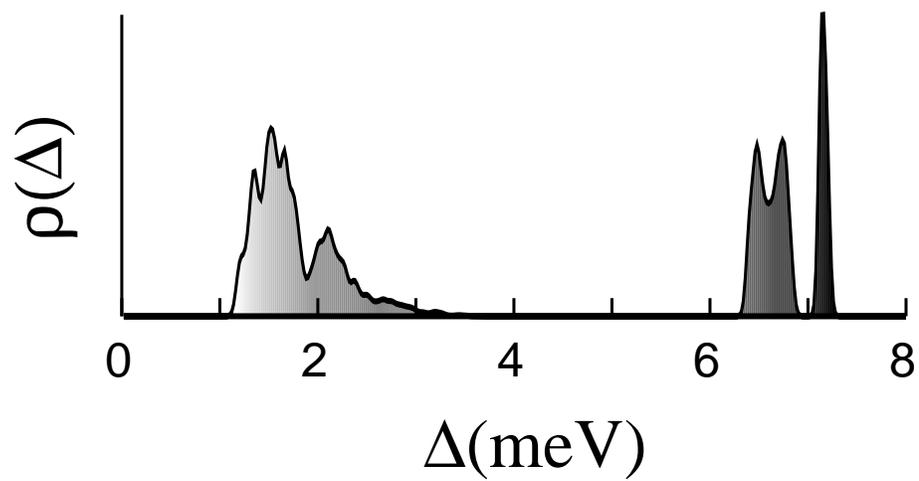}
\end{center}
\caption{The energy distribution of the gap values calculated in Ref.
\cite{Choi}.}
\label{Choi2}
\end{figure}

As is shown in Fig. \ref{Choi1}, the Fermi surface of MgB$_2$
consists of two nearly cylindrical sheets, "2D $\sigma$-bands", and
two others corresponding to the "3D $\pi$-bands". The gap on the
four Fermi surface sheets of this material has two sharp maxima:
$\Delta_1\approx 1.7\,$meV at the two   $\pi$-bands  and
$\Delta_2\approx 7\,$meV at the two $\sigma$-bands, see Fig. \ref{Choi2}.
Within each of these groups, the spread of the gap values is small, and
the gaps can be considered as constants, the ratio of which is nearly $T$
independent. In this situation, a model with only two Fermi surface
sheets and two gaps may prove  useful in relating
various macroscopic properties of MgB$_2$. Starting with Ref.
\cite{SMW}, the two-band models were studied by many, see, e.g.,
Ref. \cite{mazin3} and references therein.  We outline below the
results of evaluation of the London penetration depth and the upper
critical fields within a relatively simple two-gap weak-coupling model
which provides consistent physical picture of superconducting
anisotropies. We also discuss  possible consequences of different
anisotropies of $\lambda$ and $H_{c2}$ upon macroscopic magnetic
properties of MgB$_2$.

\subsection{Two gap model}

We begin with the notion that within the weak-coupling BSC theory the
macroscopic characteristics of superconductors such as the coherence
length and the penetration depth are expressed in terms of quantities
averaged over the Fermi surface. This  basic consequence  of
the collective nature of the superconducting condensate is formally
expressed by the self-consistency (or gap) equation:
\begin{equation}
\Delta({\bf r},{\bf v},T)=2\pi TN(0) \sum_{\omega >0}^{\omega_D} \langle
V({\bf v},{\bf v}^{\prime\,}) f({\bf r},{\bf
v}^{\prime\,},\omega)\rangle_{{\bf v}^{\prime\,}}\,.   \label{eil4}
\end{equation}
Here,  $N(0)$ is the total density of states at the Fermi level per one
spin; the Fermi velocity ${\bf v}$  relates to a certain position on
the Fermi surface, $\Delta$ is the gap function,
$f({\bf r},{\bf v},\omega)$ is  Eilenberger Green's
function which describes the condensate, and $\hbar\omega=\pi
T(2n+1)$ with an integer
$n$. In the uniform zero-field situation
$f=\Delta/\sqrt{\Delta^2+\hbar^2\omega^2}$. Substituting this
in Eq. (\ref{eil4}) one can evaluate how the gap depends on
temperature and on position at the Fermi surface provided the
pair-pair interaction $V({\bf v},{\bf v}^{\prime\,})$ is known.

The averages over the Fermi surface weighted with the local
density of states $
\propto 1/  v $ are defined as
\begin{equation}
\langle X \rangle = \int \frac{d^2{\bf k}_F}{(2\pi)^3\hbar N(0)  v}\,
\,X({\bf v})\,. \label{<>}
\end{equation}

Commonly, the interaction $V$ is assumed factorizable \cite{Kad},
  $ V({\bf v},{\bf v}^{\prime\,})=V_0 \,\Omega({\bf v})\,\Omega({\bf
v}^{\prime\,})$ (the assumption with far reaching consequences for the
two-gap model, see below). Then, one looks for
$\Delta ( {\bf r},T;{\bf v})=\Psi ({\bf r},T)\, \Omega({\bf v})$. The
self-consistency Eq. (\ref{eil4}) for the uniform condensate takes the
form:
  \begin{equation}
1=2\pi T N(0)V_0 \sum_{\omega >0}^{\omega_D}\Big\langle
\frac{\Omega^2({\bf v} )  }{\sqrt{\Psi^2\Omega^2({\bf
v})+\hbar^2\omega^2}}\Big\rangle \,.
\label{gap}
\end{equation}

   Consider now a model material with the gap anisotropy given by
\begin{equation}
\Omega ({\bf v})= \Omega_{1,2}\,,\quad {\bf v}\in   F_{1,2} \,,
  \label{e50}
\end{equation}
where $F_1,F_2$ are two sheets of the Fermi surface.  Denoting
the densities of states on the two parts as $N_{1,2}$, and assuming the
quantity $X$ being constant at each sheet, we obtain for the general
averaging:
\begin{equation}
\langle X \rangle = \frac{\langle X \rangle_1 N_1+\langle X \rangle_2
N_2}{N(0)} =
\nu_1\langle X
\rangle_1+
\nu_2\langle X \rangle_2\,,\label{norm2}
\end{equation}
where we use normalized densities of state
$\nu_{1,2}= N_{1,2}/N(0)$ for brevity.  The Fermi relief of the gap,
$\Omega$, can be normalized so that $\langle \Omega^2 \rangle=1$ which
gives for the two-gap model:
\begin{equation}
\Omega_1^2 \nu_1+\Omega_2^2\nu_2=1\,,\quad \nu_1+\nu_2=1\,.\label{norm1}
\end{equation}

Within this model, the interaction $V$ now takes the form of a $2\times 2$
symmetric matrix $V_{ij}/V_0$ with diagonal components $\Omega_1^2$
and $\Omega_2^2$, and the two equal off-diagonal elements $\Omega_1\Omega_2
$. It is worth noting that if $V_{12}=0$, i.e., there is no interband
interaction at all, the general gap equation (\ref{eil4}) splits in two
independent equations for the independent order parameters, $\Delta_1$
on $F_1$ and $\Delta_2$ on $F_2$. Such a model would imply different,
in general, critical temperatures for $\Delta$'s, different phases
with many macroscopic consequences \cite{Askerzade,Gurevich}.
The factorazable interaction $V_{ij}=V_0\Omega_i\Omega_j$
excludes such possibilities and automatically implies the single
$T_c$ for both gaps, the feature in fact seen in many tunneling experiments
on MgB$_2$ \cite{Giubileo,tunnel1,tunnel2}.

\subsection{${\bf T}$ dependence of the two gaps}

The self consistency equation (\ref{gap}) for the order parameter
$\Psi(T)$ can be solved for known $\Omega_{1,2}$ and the densities of
states $\nu_{1,2}$. Based on the band structure calculations, the relative
  densities of states
$\nu_1$ and $\nu_2$ of our model are $\approx\,$0.56 and 0.44
\cite{Bel,Choi}. The ratio $\Delta_2/\Delta_1=\Omega_2/\Omega_1\approx
4$, if one takes the averages of $6.8\,$ and $1.7\,$meV
for the two groups of distributed gaps as calculated in Ref.
\cite{Choi}. Then, the normalization (\ref{norm1})  yields
$\Omega_1=0.36$  and  $\Omega_2=1.45$. It should be noted that the
tunneling data provide somewhat lower ratio
$\Delta_2/\Delta_1$ \cite{Giubileo,tunnel1,tunnel2}.

\begin{figure}
\begin{center}
\includegraphics*[width=12cm]{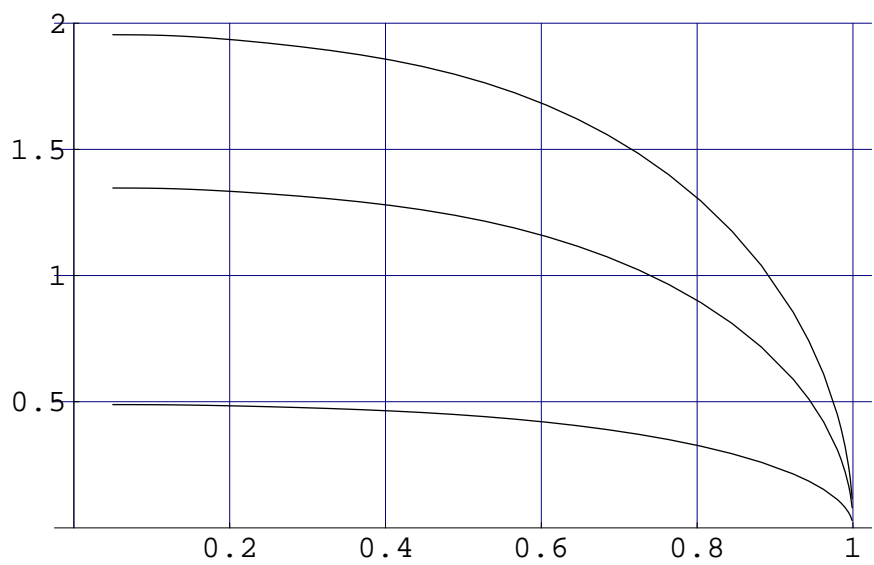}
\end{center}
\caption{The  gaps $\Delta_{1,2}=\Psi(T)\,\Omega_{1,2}$ versus
$T/T_c$. The upper
curve is $\Delta_2/T_c$, the lower one is $\Delta_1/T_c$, and the
middle curve is
$\Psi(T)/T_c$ evaluated as described in the text.}
\label{figg1}
\end{figure}

We have now all  parameters needed to solve the self-consistency equation
  (\ref{gap}) for $\Psi(T)$ (the clean case). This
is done numerically and the result   is shown in Fig.
\ref{figg1} along with two gaps $\Delta_{1,2}(T)$.

The parameter $\Delta/T_c$ commonly used to distinguish strong and
weak  coupling materials, has a "strong coupling" value of $\approx 4$
for $\Delta_2/T_c$, whereas $\Delta_1/T_c\approx 1$, which is less than
the   BCS  weak coupling value of 3.5 . Thus, this parameter cannot be
used to characterize coupling for materials like MgB$_2$.
  The $T$ dependence of the two gaps similar to shown in Fig.
\ref{fig1} has been reported by a number of groups 
\cite{Giubileo,tunnel1,tunnel2}.

\subsection{Anisotropy of ${\bf\lambda}$}

Having obtained the temperature dependence of both gaps, one can turn to
   calculation of the weak-field penetration depth $\lambda (T)$. To this
end, one first solves equations of superconductivity for the
zero-field case and then employs the perturbation theory to evaluate the
small corrections due to small currents. In this manner one obtains the
London relation between the current and the vector potential,
\begin{equation}
{4\pi\over c} j_i =-  (\lambda^2)_{ik}^{-1}
\Big({2\pi\over\phi_0} {\partial\theta\over\partial x_k} + A_k \Big)
\end{equation}
where $\theta$ is the phase, $\phi_0$ is the flux quantum, $\bf A$ is the
vector potential, and
\begin{equation}
(\lambda^2)_{ik}^{-1}= \frac{16\pi^2 e^2T}{
c^2}\,N(0)  \sum_{\omega} \Big\langle\frac{
\Delta^2v_iv_k}{\beta ^{3}}\Big\rangle \,  \label{e31}
\end{equation}
  is the tensor of the inverse squared penetration depth.
Here, $i,k=x,y,z$ and $\beta ^2=\Delta^2+ \hbar^2\omega^2$
(the reader may find details of this calculation in Ref.
\cite{KRap}).  This general expression for the tensor of
squared penetration depth is valid for clean superconductors with an
arbitrary Fermi surface and gap anisotropies.

The anisotropy parameter now reads:
\begin{equation}
\gamma_{\lambda}^2=\frac{\lambda^2_{cc}}{\lambda^2_{aa}}= \frac{
  \langle  v_a^2\,\Delta_0^2\sum_{\omega}\beta ^{-3}\rangle
}{ \langle
v_c^2\,\Delta_0^2\sum_{\omega}\beta ^{-3}\rangle }
  \,. \label{gamma(T)}
\end{equation}

As $T\to 0$, we have $2\pi
T\Delta_0^2 \sum_{\omega}\beta ^{-3}\to 1$, and
\begin{equation}
\gamma_{\lambda}^2(0)=  \frac{  \langle v_a^2\rangle}{\langle
v_c^2\rangle}\,.
\label{gamma(0)}
\end{equation}
Note that the gap and its anisotropy do not enter this result. The physical
reason for this is in the Galilean invariance of the superfluid flow in the
absence of scattering: all charged particles take part in the
supercurrent.

  Near $T_c$, the sum over $\omega$ in Eq. (\ref{gamma(T)}) is is a ${\bf
k}_F$ independent constant because $\beta\approx\hbar\omega$, and  we
obtain the GL result of Ref. \cite{Gorkov}:
\begin{equation}
\gamma_{\lambda} ^2(T_c) = \frac{\lambda^2_{cc}}{\lambda^2_{aa}}=
\frac{\langle \Omega^2 v_a^2\rangle }{\langle \Omega^2 v_c^2\rangle}\,.
\label{anis_clean}
\end{equation}
   We observe that
near $T_c$, the gap anisotropy amplifies contribution of
  the large gap Fermi surface pieces  to the anisotropy parameter
$\gamma_{\lambda}$. We also see that for isotropic gaps,
$\gamma_{\lambda}(T_c)=\gamma_{\lambda} (0)$ (which is not true for the
general case).  Thus,  the anisotropy parameter depends on
$T$, the feature absent in  superconductors with isotropic gaps. \\

To apply these results to MgB$_2$, one has to know averages of squared
Fermi velocities not only over the whole Fermi surface (as for
$\gamma_{\lambda} (0)$ of Eq. (\ref{gamma(0)})), but also the averages
over the separate Fermi sheets. These can be taken from the band structure
calculations of Ref. \cite{Bel}:
\begin{eqnarray}
\langle v_a^2\rangle_1&=&33.2\,,\quad \langle v_c^2\rangle_1=42.2,
\nonumber\\
\langle v_a^2\rangle_2&=&23\,,\qquad
  \langle v_c^2\rangle_2=0.5\times 10^{14}\,cm^2/s^2.
\end{eqnarray}
  The  numerical results for $\gamma_{\lambda}(T)$ are shown in Fig.
\ref{figg2}.

Thus, we expect the London penetration depth of clean MgB$_2$ to be nearly
isotropic at low temperatures; $\gamma_{\lambda}(T)$ increases on heating
and reaches $\approx 2.6$ at $T_c$. Qualitatively similar behavior of
$\gamma_{\lambda}(T)$ is predicted within a more general Eliashberg scheme
   \cite{Golubov}. As up to date, there is no
published data on direct measurements of the $\lambda$ anisotropy
and its $T$ dependence. It is pointed out below that the data on
the angular dependence of the torque acting on the anisotropic
superconducting crystal in tilted fields cannot provide
information about $\gamma_{\lambda}(T)$ without taking into
account the $H_{c2}$ anisotropy which  for MgB$_2$ might be quite
different from $\gamma_{\lambda}(T)$. In principle, one can
extract $\gamma_{\lambda}$ from the data on the vortex lattice
structure in fields tilted relative to the $c$ axis of MgB$_2$.
These, however, are difficult to acquire and to our knowledge are
not yet available.

\subsection{Anisotropy of ${\bf H_{c2}}$}

Unlike the case of London $\lambda$, the evaluation of $ H_{c2}$ is more
complicated a problem because one has to deal with the high-field phase
transition from the normal to the superconducting mixed state. Still,
major behavior of $H_{c2}(T)$ can be obtained within a simplified
scheme: Since the  actual band structure enters equations for
$H_{c2}$   via Fermi-surface averages, one can further model the sheets
$F_{1,2}$ by two spheroids with average Fermi velocities close to the
band-structure generated values. As a result one obtains qualitatively (and
- given the spread of existing data - quantitatively) correct behaviors of
$H_{c2}(T)$ for both principal directions. Details of
this evaluation  can be found in Ref. \cite{MMK}. The result for the
anisotropy parameter $\gamma_H(T)=H_{c2,ab}/H_{c2,c}$ obtained within the
scenario of two constant gaps on two Fermi surface sheets, the same used
for $\gamma_{\lambda}(T)$ and with the same input parameters, is shown in
Fig. \ref{figg2}.

The drop of the  $H_{c2}$ anisotropy with increasing $T$ has been
recorded   in measurements on single crystals of
MgB$_2$ \cite{angst}.  Nearly the same behavior has been extracted from
  the data on  random polycrystals   in the whole
temperature range using the $T$ dependence of the magnetization of
random powders \cite{BC}. Recent specific heat measurements
   \cite{Lyard} and magnetization data
\cite{Welp} on single crystals produced similar results.  All these
data show qualitatively similar behavior to that of the upper curve
of Fig. \ref{figg2}.

Physically, the large anisotropy of $H_{c2}$ at low temperatures
($\approx 6$ in our calculation) is related to the large gap value at
the Fermi sheet which is nearly two-dimensional. With increasing $T$,
the thermal mixing with the small-gap states on the three-dimensional
Fermi sheet suppresses the anisotropy down to 2.6 at $T_c$.

\subsection{ Free energy and torque in materials with
different anisotropies of ${\bf\lambda}$ and ${\bf H_{c2}}$}

  One of the most sensitive methods used to extract the anisotropy
parameter $\gamma$ is to measure the torque acting on a
superconducting crystal in the mixed state with the applied
field tilted relative to the crystal axes.  In intermediate
field domain, $H_{c1}\ll H\ll H_{c2}$, the demagnetization shape effects
are weak, and the torque density can be evaluated: \cite{K88,farr}
\begin{equation}
\tau = \frac{\phi_0B(\gamma^2-1)\sin
2\theta}{64\pi^2\lambda^2\gamma^{1/3}
\varepsilon(\theta)}\, \ln\frac{\eta\,H_{c2,a} } { B
\varepsilon(\theta) } \,,
\label{trq_0}
\end{equation}
where $\theta$ is the angle between the induction $\bf B$ and the
crystal axis $c$,
\begin{equation}
\varepsilon(\theta) =\sqrt{\sin^2\theta+\gamma^2 \cos^2\theta}\,,
\label{eps}
\end{equation}
$\lambda^3=\lambda_{a}^2\lambda_c$, and $\eta\sim 1$. This formula
can be written as ${\bf \tau}={\bf M}\times{\bf H}$; since in this
field domain the magnetization $M\ll H$, one can disregard the
distinction between ${\bf B}$ and ${\bf H}$.  It has been assumed in
the derivation of Eq. (\ref{trq_0})  that the anisotropies of
$H_{c2}$ and of the London penetration depth are the same: $\gamma_H
=\gamma_{\lambda} =\gamma$.
It is worth noting that Eq. (\ref{trq_0}) describes the system with the
stable equilibrium at $\theta=\pi/2$, i.e., the uniaxial crystal in the
external field  positions itself so that the   field is parallel to
$ab$ (if $\gamma > 1$).

Expression (\ref{trq_0}) for the torque has been derived within the
London approach by employing the cutoff at distances on the order of
the coherence length $\xi$ where this approach fails. The formula,
however, has been confirmed experimentally (for a few high-$T_c$ materials
close to their $T_c$) with a good accuracy as far as the angular
dependencies of quantities involved are concerned \cite{Far2}.
Uncertainties of the London approach are incorporated in the parameter
$\eta\sim 1$; discussion of those can be found, e.g., in Ref.
\cite{nonloc}.

We now outline possible effects of different anisotropies of $\lambda$ and
$H_{c2}$ on the torque angular dependence. To this end, we  first recall
the London expression for the free energy valid for intermediate fields
$H_{c1}\ll H\ll H_{c2}$   along
$z$  tilted with respect to the $c$ crystal axis over the angle $\theta$
toward the crystal direction $a$ \cite{CDK}:
\begin{equation}
F=\frac{B^2}{8\pi }+\frac{B^2m_{zz}}{8\pi\lambda^2m_a}\, \sum_{\bf
G}\frac{1}{m_{zz}G_x^2+m_cG_y^2}\,;
\label{energy}
\end{equation}
here, $m_{zz}=m_a\sin^2\theta+m_c\cos^2\theta $,
$m_c/m_a=\gamma_{\lambda}^2\,$, $m_a^2m_c=1$ for uniaxial crystals, and
${\bf G}$ form the reciprocal vortex lattice. The summation is
extended over all nonzero ${\bf G}$.

For isotropic case, all reduced masses in Eq.(\ref{energy}) are unity.
The sum $\sum G^{-2}$ can be approximated by an integral
$(\phi_0/4\pi^2 B)\int dG_xdG_y/G^2 $ which diverges logarithmically and
should be cut off at a circle in the reciprocal space of a radius $\sim
1/\xi$ where $\xi$ is the radius of the circular vortex core.

In anisotropic material with $\gamma_{\lambda}=\gamma_H$, calculation of
this sum is reduced to the isotropic case   by rescaling
\begin{equation}
g_x=\sqrt{m_{zz}}\,G_x\,,\qquad g_y=\sqrt{m_c}\,G_y\,. \label{scaling}
\end{equation}
In this manner one
can derive the free energy of a superconductor in an applied field of a
certain direction; the torque formula (\ref{trq_0}) is obtain
differentiating the energy with respect to the angle $\theta$.

However, for $\gamma_{\lambda}\ne \gamma_H$, the masses in the sum
(\ref{energy}) correspond to the anisotropy $\gamma_{\lambda}$ (in other
words, the vortex-vortex interaction is determined by the $\lambda$
anisotropy) whereas the cut-off reflects the shape of the core (it is
determined by the anisotropy of the coherence length $\xi$, i.e., by the
parameter
$ \gamma_H$). In this situation, the scaling (\ref{scaling}) which
transforms the summand of Eq. (\ref{energy}) to the isotropic-like form,
leaves the shape of the cut-off contour elliptical. Therefore, the energy
$F$ and its angular dependence acquire extra terms and so does the torque
which can be measured.

We refer the reader to the original work \cite{trq2002} for the details of
this calculation and give here only the result.
To write explicitly the angular dependence of $F$, it is convenient to
use the angular functions
\begin{equation}
\Theta_{\lambda,H}(\theta)=\varepsilon_{\lambda,H}(\theta)/\gamma_{\lambda,H}
\end{equation}
where   $\varepsilon_{\lambda,H}(\theta)$ are defined in Eq.
(\ref{eps}) with corresponding $\gamma$'s.
In terms of these functions,
\begin{equation}
  F=\frac{B^2}{8\pi}+\frac{\phi_0B
\Theta_{\lambda}}{32\pi^2\lambda_{ab}^2}\,
  \ln\frac{2\sqrt{3}\,\mu_a\phi_0\Theta_{\lambda
}}{\xi^2B \,(\Theta_{\lambda }+\Theta_{H })^2} \,,
\label{FF}
\end{equation}

where $\mu_a = \gamma_H^{-2/3}$.
The torque density $\tau = -\partial F/\partial \theta$ follows:

\begin{eqnarray}
\tau  &=&
\frac{\phi_0B(\gamma_{\lambda}^2-1)\sin 2\theta}
{64\pi^2\lambda^2\gamma_{\lambda}^{4/3}\Theta_{\lambda}}\,
\Big[\ln\Big(\frac{\eta H_{c2,c}}{B}\,\frac{4e^2
\Theta_{\lambda}
  }{(\Theta_{\lambda }+\Theta_{H })^2}\Big)\nonumber\\
&-&
\frac{2
\Theta_{\lambda}}
{\Theta_{\lambda}+\Theta_H}\Big(1+\frac{\Theta_H^{\prime}}
{\Theta_{\lambda}^{\prime}}\Big)\Big]\,.
\label{trq1}
\end{eqnarray}
where $e=2.718...\,$.
  In the standard case of $\gamma_H=\gamma_{\lambda}=\gamma$, Eq.
(\ref{trq1}) reduces to the  result (\ref{trq_0}).

Since both $\Theta_{\lambda,H}^{\prime} < 0$, the
second  contribution to the torque (\ref{trq1}) is negative whereas the
first one is positive. The positive torque implies that the system
energy decreases with increasing $\theta$, as in the standard case of
$\gamma_H=\gamma_{\lambda}$ for which   $\theta=\pi/2$ is the
stable equilibrium.

On the other hand, it is straightforward to see that for large enough
  $\gamma_H$ and fixed  $ \gamma_{\lambda} $ (e.g., set $\gamma_{\lambda}=1
$ and leave $\gamma_H>1$) Eq. (\ref{trq1}) yields a negative torque,
which means that in the equilibrium   the crystal $c$ axis is parallel to
the applied field. This strange behavior can be tracked down to the
angular dependence of the line energy of a single vortex:  the energy cost
of having an elliptical core for vortices along $ab$ for which the current
distribution far from the core is near isotropic (as for the example with
$\gamma_{\lambda}=1$) is too high.

To illustrate how the angular dependence of the torque varies with
  anisotropies of $H_{c2}$ and  $\lambda$,  we evaluate numerically
the torque density (\ref{trq1}) for parameters in the range of those
for clean MgB$_2$.  Fig. \ref{fig1} shows $\tau(\theta)$ for
$\gamma_{\lambda}=2.2$ and $\gamma_H=3$, the values expected for
temperatures somewhat below $T_c$. Qualitatively, the
dependence is standard; the torque is positive in the whole angular
domain, i.e., $\theta=\pi/2$ is the stable equilibrium.

With decreasing $T$, $\gamma_H$ of MgB$_2$ increases whereas
$\gamma_{\lambda}$ decreases. In Fig. \ref{fig2} the torque
(\ref{trq1}) is plotted for $\gamma_{\lambda}=2$, $\gamma_H=5$
(the upper curve) and for $\gamma_{\lambda}=1.7$, $\gamma_H=5.3$
(the lower curve). These values
correspond roughly to  0.7 and 0.6$\, T_c$ according to Ref.
\cite{MMK}. Clearly, $\theta=\pi/2$ as well as
$\theta=0$ are unstable; the stable equilibrium is shifted to
$0<\theta<\pi/2$.

Finally, we plot in Fig. \ref{fig3} the torque density for parameters
which correspond to low temperatures, where $\gamma_{\lambda}\approx 1.1 $
and $\gamma_H\approx 6$. The torque is negative for all angles
implying  the stable equilibrium at   $\theta=0$.

\section{Conclusions}

Hence, different gaps at different Fermi surface pieces of MgB$_2$
(or generally, anisotropic gaps on anisotropic Fermi surfaces) may
lead to profound macroscopic consequences such as those considered
above. Moreover, since $\gamma_H$ determines the anisotropy of the
coherence length and, therefore, of the vortex core, whereas
$\gamma_{\lambda}$ describes the ellipticity of the current distribution
far from the core, the  symmetry of the intervortex interaction should
depend on the intervortex spacing, i.e., on the field $B$ and its
direction.  In clean MgB$_2$ at low temperatures in fields along $ab$, one
expects to have the standard triangular (hexagonal) vortex lattice in low
fields ($\gamma_{\lambda}\sim 1$), which should evolve to a distorted
triangular (orthorhombic) lattice in increasing fields when the core
ellipticity ($\gamma_H\approx 6$) becomes relevant for the vortex
current distribution. The field dependence of the vortex lattice
structure should become weaker at elevated temperatures and disappear
near $T_c$ where $\gamma_{\lambda}\to \gamma_H$. In intermediate
temperature range where the equilibrium vortex lattice orientation shifts
from
$\theta=\pi/2$ to lower angles,
  one may expect peculiar vortex dynamics (as for fields near parallel to
the layers of high-$T_c$ materials). This possibility  calls for further study.

\begin{figure}
\begin{center}
\includegraphics*[width=12cm]{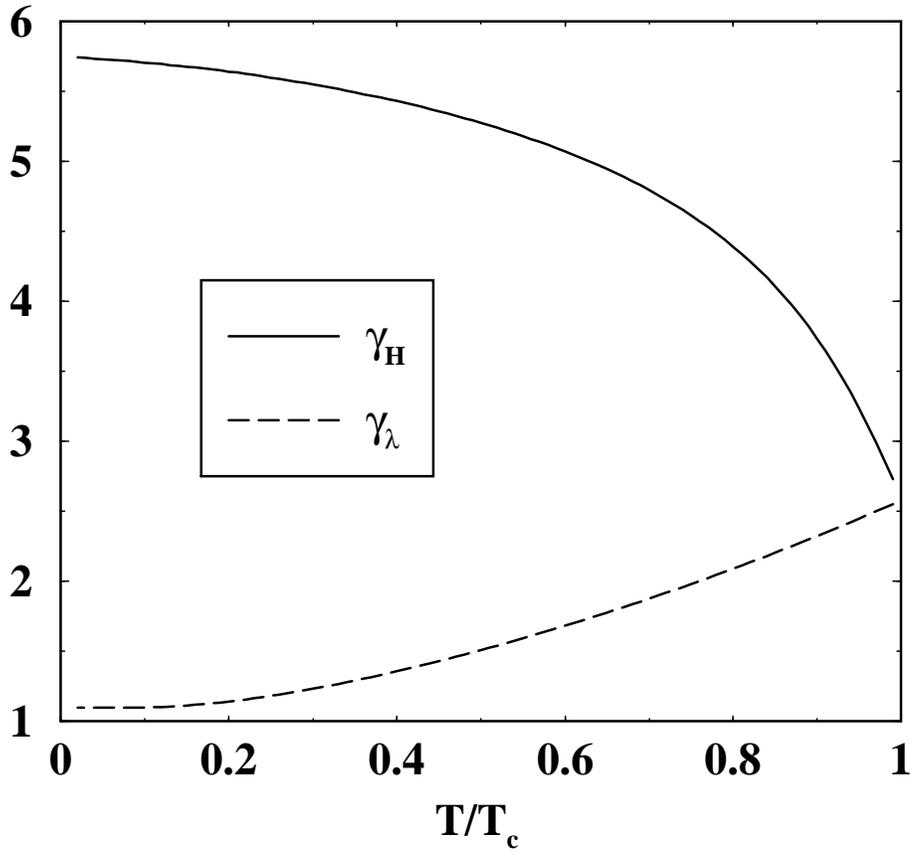}
\end{center}
\caption{Anisotropy ratio $\gamma_H=H_{c2,ab}/H_{c2,c}$ versus
$ T/T_c$ calculated with parameters for MgB$_2$ given in the text.
Dashed line is $\gamma_\lambda(T)=\lambda_c/\lambda_{ab}$.}
\label{figg2}
\end{figure}

\begin{figure}
\begin{center}
\includegraphics*[width=12cm]{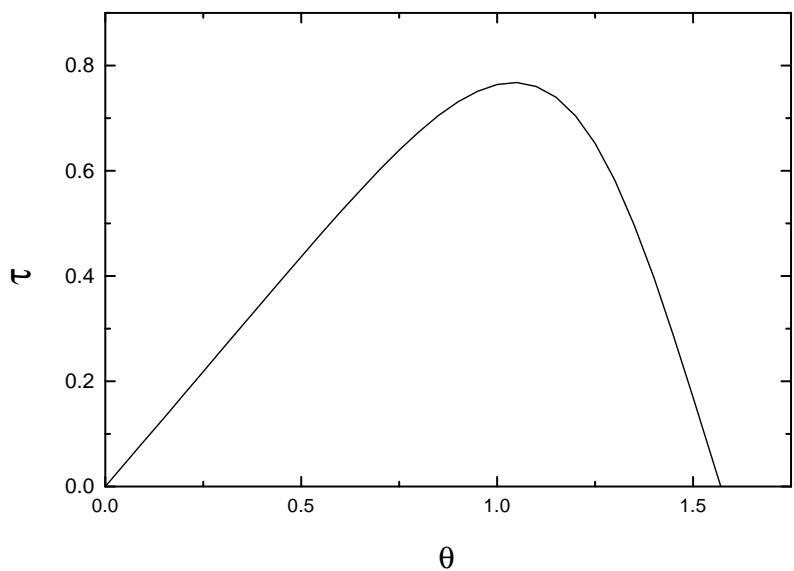}
\end{center}
\caption{The torque $\tau$ in units of
$\phi_0B /32\pi^2\lambda_{ab}^2$ {\it versus} angle
$0<\theta<\pi/2$ for $\gamma_{\lambda}=2.2$, $\gamma_H=3$, and
$4e^2\eta H_{c2,c} /B=100$.}
\label{fig1}
\end{figure}

\begin{figure}
\begin{center}
\includegraphics*[width=12cm]{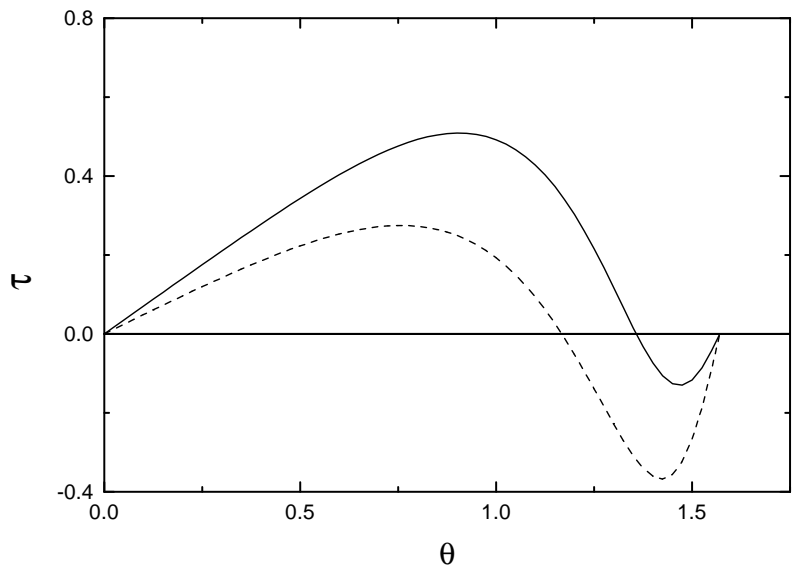}
\end{center}
\caption{The same as in Fig.\ref{fig1}. The
solid curve is calculated with Eq. (\ref{trq1}) for
$\gamma_{\lambda}=2$ and $\gamma_H=5$;  the dashed curve corresponds to 
$\gamma_{\lambda}=1.7$ and $\gamma_H=5.3\,$.}
\label{fig2}
\end{figure}

\begin{figure}
\begin{center}
\includegraphics*[width=12cm]{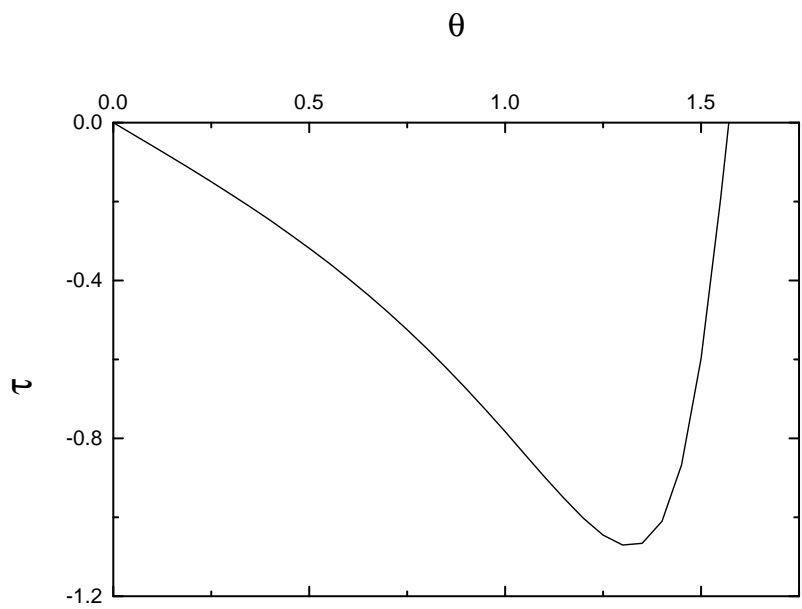}
\end{center}
\caption{The same as in Fig.\ref{fig1}, but $\gamma_{\lambda}=1.1$
and $\gamma_H=6$.}
\label{fig3}
\end{figure}

Thus, the question ``what is the anisotropy parameter of MgB$_2$?" does
not have a unique answer. To pose the question properly one should
specify the quantity of interest. If this is $H_{c2}$, the answer is
given by the upper curve of Fig. \ref{figg2} for a clean material; if this
is
$\lambda$, see the dashed line. If this is the magnetization in
intermediate fields, $M\propto (\phi_0/\lambda^2)\ln (H_{c2}/H)$, the
main contribution to anisotropy comes from $\lambda$; however, the
$H_{c2}$ anisotropy contributes as well (being smoothed by the
logarithm). The last situation should be taken into account while
extracting anisotropy from the  torque data in tilted fields
\cite{angst,torque1}. It should be noted in closing, that all
anisotropies discussed here might be suppressed by impurities.

The authors are grateful to P. Canfield for numerous and
   helpful discussions.  Ames Laboratory is operated for the U. S.
Department of Energy by Iowa State University under contract No.
W-7405-ENG-82.  The work was supported by the Director of Energy Research,
Office of Basic Energy  Sciences, U. S. Department of Energy.

\clearpage

\clearpage

\clearpage

\clearpage

\clearpage

\clearpage

\clearpage

\clearpage

\clearpage

\clearpage

\end{document}